\documentclass[a4paper]{article}

\usepackage{INTERSPEECH2021}
\usepackage{algorithm}
\usepackage{algorithmic}
\usepackage{subfigure}
\usepackage{enumitem}

\title{Dropout Regularization for Self-Supervised Learning of Transformer Encoder Speech Representation}
\name{Jian Luo, Jianzong Wang*\thanks{*Corresponding author: Jianzong Wang, jzwang@188.com}, Ning Cheng, Jing Xiao}
\address{Ping An Technology (Shenzhen) Co., Ltd.}

\begin{document}

\maketitle
\begin{abstract}
Predicting the altered acoustic frames is an effective way of self-supervised learning for speech representation. However, it is challenging to prevent the pretrained model from overfitting. In this paper, we proposed to introduce two dropout regularization methods into the pretraining of transformer encoder: (1) attention dropout, (2) layer dropout. Both of the two dropout methods encourage the model to utilize global speech information, and avoid just copying local spectrum features when reconstructing the masked frames. We evaluated the proposed methods on phoneme classification and speaker recognition tasks. The experiments demonstrate that our dropout approaches achieve competitive results, and improve the performance of classification accuracy on downstream tasks.
\end{abstract}

\section{Introduction}
In recent years, deep-learning models have shown remarkable success in speech tasks, such as automatic speech recognition, speaker identification, spoken language understanding, etc~\cite{Gulati2020Conformer, Luo2021Unidirectional, Ding2020AutoSpeech, qin2021cointeractive}. Among these models, transformer-based architectures have obtained a substantial performance improvement. Despite these achievements, the collection of paired speech data still confuses researchers and engineers. Speech data requires intensive labeling and aligning works which can only be done manually. On the contrary, unpaired speech data are much more available than paired ones. With hardly any data collection cost, it seems to be an appealing solution for the dilemma that the supervised learning is facing. Therefore, the research community is shifting its focus to self-supervised or semi-supervised learning~\cite{baskar2019semisupervised,fan2020unsupervised,Karita2018Semi,Hori2019Cycle}.

Self-Supervised Learning (SSL) is an approach of learning data representation from unlabeled data, and retraining the model on labeled data~\cite{baevski2020wav2vec}. In this paper, we focus on the SSL of transformer network, to extract high-level speech representation. Through SSL pretraining, learned transformer models could be applied to downstream Speech and Language Processing (SLP) tasks. Recent works have proposed several kinds of SSL schemes. Autoregressive Predictive Coding (APC)~\cite{chung2019unsupervised} and Contrastive Predictive Coding (CPC)~\cite{Oord2018Representation} focus on maximizing the probability of predicting future frames and the contrastive loss from separating negative sample set respectively. APC and CPC are based on unidirectional RNN, which limits speech representation learning without attending future frames. \cite{Wang2020Unsupervised} proposed to use bidirectional RNN in the pretraining, and incorporated it to bidirectional speech recognition models. Audio Word2Vec~\cite{chung2016audio} generates vector representation for audio segments, which is trained by an RNN-based autoencoder, to reconstruct the input speech audio. VQ-wav2vec\cite{baevski2020vqwav2vec} learns discrete speech representation of audio segments using VQ-VAE style codebook. VQ-wav2vec takes discrete tokens as the input, and achieves impressive results on speech recognition tasks. 

Transformer-based SSL models use multi-layer transformer encoder to predict masked frames or spectrum bands, forcing the model to learn hidden speech features from both directions. The learned features could serve as the input of downstream SLP tasks. Masked Predictive Coding (MPC)~\cite{jiang2019improving} applied transformer-based model to unsupervised pretraining, improving the performance of speech recognition task. Mockingjay~\cite{Liu2020Mocking} introduced BERT-style masking strategy into the pretraining of speech representation. TERA~\cite{liu2020tera} proposed to use multi-target auxiliary task to pretrain the transformer encoder, handling alteration on temporal, channel, and magnitude axes. Speech XLNet\cite{Song2020} presented an XLNet-style pretraining scheme, introducing dynamic permutation for further exploitation of the speech. Despite their outstanding performance on downstream tasks, above mentioned models may still suffer from overfitting or degradation issues. The models may just exploit the local smoothness of the speech, simply copying local spectrum features when reconstructing the masked frames. To solve this issue, APC proposed to predict frames that are several steps away\cite{chung2019unsupervised}. MPC and TERA mask a consecutive range of speech frames instead of a single frame in temporal axis. In this work, we proposed to introduce dropout regularization to alleviate this problem.

Dropout~\cite{hinton2012improving, Srivastava2014Dropout} is a popular method of model regularization for fully-connected neural network. It forces the network to discard some neural units randomly, and learn new pattern using available connections and parameters~\cite{chen2020selectscale}. Dropout has also been applied in many deep learning network structures, including RNN~\cite{Watt2018Dropout} and CNN~\cite{cai2020effective}. Dropout was used in many tasks to prevent model from overfitting. In natural language processing, \cite{zhang2020token} proposed to use dropout in machine translation, making the model to generate same output with less input. In the \cite{wu2020generating}, dropout is used to generate multiple translations that share similar meanings. In terms of model robustness, dropout can be applied to ensure the safety of systems and make them robust to perturbations~\cite{goodfellow2015explaining,jayashankar2020detecting}. 

Different from common dropout which cuts off the co-adaptation between units randomly, some works instead used dropout to discard the most discriminative activation regions. In weak supervised object detection, dropout is added to utilize less significant patterns and avoid overfitting the ground truth bounding box\cite{gao2020cascade,Junsuk2019Attention}. In text classification, DropAttention~\cite{Zehui2019DropAttention} regularizes the attention weights in transformer, helping the model to utilize more contextualized word vectors. In this paper, we proposed to introduce \textbf{attention dropout} and \textbf{layer dropout} to the SSL of transformer encoder speech representation. Both of the two dropout methods prevent the transformer encoder from degrading to trivial solution that copies local features. Attention dropout and layer dropout encourage the model to use the features that are far apart from current predicting frame, hence capturing global speech information.

\vfill\pagebreak

\section{Proposed Method}
In this paper, we proposed to use dropout regularization for the SSL of speech. The architecture is based on TERA (Transformer Encoder Representations from Alteration)~\cite{liu2020tera}, which pretrains the model with three auxiliary objectives: (1) time alteration, (2) channel alteration, (3) magnitude alteration. We introduce two dropout methods into the transformer encoder: (1) attention dropout, reconstructing the attention weight matrix of self-attention mechanism, (2) layer dropout, masking the most active elements of feed-forward layer. The transformer encoder network is pretrained by reconstructing the altered acoustic features. After that, the hidden states of last layer are extracted, and will be incorporated to downstream tasks.

\begin{figure}[ht]
	\begin{center}
		\centerline{\includegraphics[width=0.85\columnwidth]{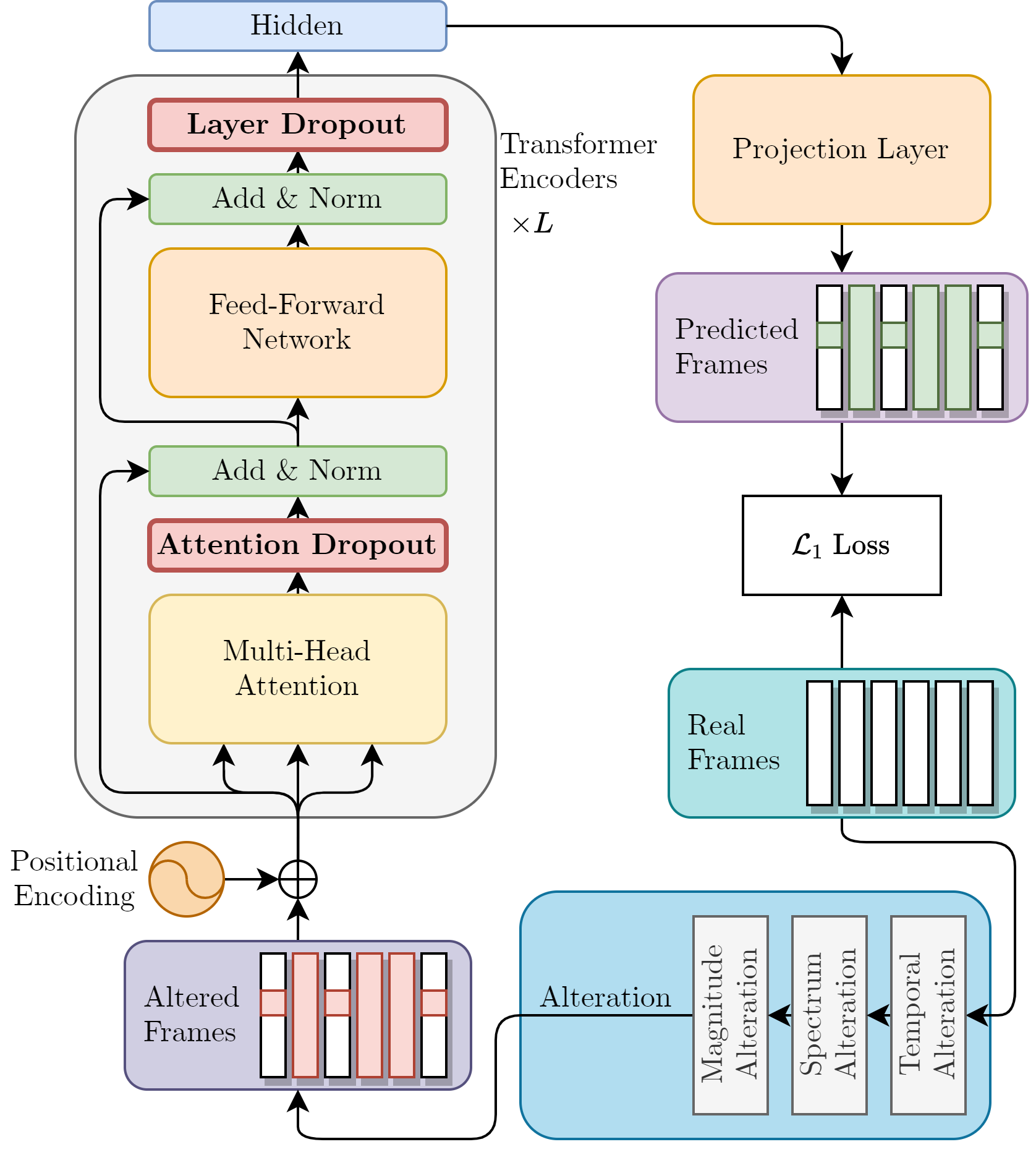}}
		\caption{The Transformer Encoder Architecture of Self-Supervised Learning with dropout regularization}
		\label{fig1}
	\end{center}
\end{figure}

\subsection{Architecture}
Transformer~\cite{Vaswani2017Attention} has impressive performance in the SSL of speech representation. Our model architecture uses a multi-layer transformer encoder with multi-head self-attention mechanism, illustrated in Figure~\ref{fig1}. The input audio sequence $\mathbf{X}\in T\times D_{mel}$ are fed into the network, where $T$ is the audio frames length, and $D_{mel}$ is the dimension of mel-scale features. Each encoder layer has two sub-layers: (1) a multi-head self-attention network, (2) a feed-forward layer. We apply the attention dropout $\theta_{attn}$ in self-attention network, and use the layer dropout $\theta_{layer}$ in feed-forward layer. The total amount of encoder layers are denoted as $L$, and the output of each layer $l$ is $\mathbf{X_l}$. The last layer $\mathbf{X_L}$ is projected to the reconstructed features $\mathbf{\tilde{X}}$. The model is pretrained by directly optimizing L1 loss between the input sequence $\mathbf{X}$ and output sequence $\mathbf{\tilde{X}}$:
\begin{equation}
\mathcal{L}_1 = |\mathbf{X}-\mathbf{\tilde{X}}|
\end{equation}

\subsection{Attention Dropout}
For each transformer encoder layer $l$, the input feature sequence is $\mathbf{X_{l-1}}\in T\times D_{attn}$, where $D_{attn}$ is the dimension of self-attention mechanism. The multi-head self-attention mechanism projects $\mathbf{X_{l-1}}$ into three matrices: the query matrix $\mathbf{Q_h}$, the key matrix $\mathbf{K_h}$, and the value matrix $\mathbf{V_h}$.
\begin{equation}
\mathbf{Q_h},\mathbf{K_h},\mathbf{V_h}=\mathbf{X_{l-1}}\mathbf{W_h^Q},\mathbf{X_{l-1}}\mathbf{W_h^K},\mathbf{X_{l-1}}\mathbf{W_h^V}
\end{equation}
\begin{equation}
\mathrm{Attn}(\mathbf{Q_h},\mathbf{K_h},\mathbf{V_h})=\mathbf{A_h}\mathbf{V_h}=\mathrm{softmax}\left(\frac{\mathbf{Q_h}\mathbf{K_h^T}}{\sqrt{D_{attn}}}\right)\mathbf{V_h}
\end{equation}
In which, $\mathbf{W_h^Q},\mathbf{W_h^K},\mathbf{W_h^V}$ are learnable parameters of head $h$, and $h\in [1,H]$. $\mathbf{A_h}$ is denoted as attention weight matrix. As illustrated in Algorithm~\ref{algorithm1}, the attention dropout method attempts to reweight the matrix $\mathbf{A_h}$ with probability $p_{attn}$. At first, the algorithm gets maximum value $a_h^{max}$ of weight matrix $\mathbf{A_h}$ by global max-pooling operation. Then, the attention dropout $\theta_{attn}$ is applied on each element $a_h^{ij}$ as following:  
\begin{equation}
\theta_{attn}(a_h^{ij}) = \begin{cases} 0, & \mbox{if}~a_h^{ij}>\lambda_{attn}a_h^{max} \\
a_h^{ij}, & \mbox{otherwise}\end{cases}
\end{equation}
We set a threshold ratio $\lambda_{attn}\in[0,1]$. $\theta_{attn}$ erases the high attentive locations, avoiding the model from overfitting local features. After element-wise dropout, each row vector $\mathbf{A_h^j}$ is renormalized, to ensure that the sum of attention weights remains $1$. Through attention renormalization, the multi-head attention weights will be distributed over the whole spatial dimension, encouraging the model to utilize global information.

\begin{algorithm}[ht] 
	\caption{Attention Dropout Algorithm} 
	\label{algorithm1}
	\begin{algorithmic}[1]
		\STATE {\bfseries Input:} \\
		$\mathbf{A_h}$: attention weight matrix of head $h$\\
		$p_{attn}$: probability of conducting attention dropout\\
		$\lambda_{attn}$: threshold ratio of attention dropout\\
		\STATE pick a random float number $r_{attn}\in[0,1]$
		\IF {$r_{attn}>p_{attn}$}
			\RETURN $\mathbf{A_h}$
		\ENDIF
		\STATE $a_h^{max}=\mathrm{MaxPool}(\mathbf{A_h})$
		\FOR {each weight element $a_h^{ij}$ in $\mathbf{A_h}$}
			\STATE apply the attention dropout: $a_h^{ij}=\theta_{attn}(a_h^{ij})$
		\ENDFOR
		\FOR {all row vector $\mathbf{A_h^j}$}
		\STATE normalized rescale: $\mathbf{A_h^j} = \mathbf{A_h^j}/\sum_{j=1}^T\mathbf{A_h^j}$
		\ENDFOR
		\RETURN $\mathbf{A_h}$
	\end{algorithmic} 
\end{algorithm}

\begin{table*}[ht]
	\centering
	\caption{Different Configurations on Threshold Ratio, Phoneme and Speaker Classification Results on Librispeech, Accuracy (\%)}
	\scalebox{0.85} {
	\begin{tabular}{p{6.0cm}|p{0.8cm}|p{0.8cm}|p{2.3cm}|p{2.3cm}|p{2.3cm}|p{2.3cm}}
		\hline\hline
		\textbf{Pretraining Method} & $\mathbf{\lambda_{attn}}$ & $\mathbf{\lambda_{layer}}$ & \textbf{PhonemeLinear} & \textbf{Phoneme1Hidden} & \textbf{SpeakerFrame} & \textbf{SpeakerUtterance} \\
		\hline
		3L-TERA-base~\cite{liu2020tera} & -- & -- & 70.65~(65.1) & 78.51~(77.3) & 99.52~(98.9) & \textbf{99.47}~(99.2) \\
		\hline
		3L-Encoder + Attention Dropout & 0.9 & -- & 70.56 & 78.69 & 99.27 & 99.26 \\
		3L-Encoder + Attention Dropout & 0.8 & -- & 70.91 & 78.79 & 99.51 & 99.35 \\
		3L-Encoder + Attention Dropout & 0.6 & -- & 70.85 & 78.57 & 99.45 & 99.30 \\
		3L-Encoder + Attention Dropout & 0.4 & -- & 69.08 & 77.27 & 99.44 & 99.36 \\
		\hline
		3L-Encoder + Layer Drop & -- & 0.9 & 70.45 & 78.54 & 99.24 & 99.23 \\
		3L-Encoder + Layer Drop & -- & 0.8 & 71.11 & 78.72 & 99.51 & 99.33 \\
		3L-Encoder + Layer Drop & -- & 0.6 & 71.19 & 78.68 & 99.46 & 99.42 \\
		3L-Encoder + Layer Drop & -- & 0.4 & 69.07 & 76.90 & 99.21 & 98.94 \\
		\hline
		3L-Encoder + Attention \& Layer Dropout & 0.8 & 0.6 & 70.71 & 78.64 & 99.37 & 99.35 \\
		3L-Encoder + Attention \& Layer Dropout & 0.9 & 0.9 & 71.12 & 78.95 & 99.51 & 99.31 \\
		3L-Encoder + Attention then Layer Dropout & 0.8 & 0.6 & 70.88 & 78.76 & \textbf{99.52} & 99.33 \\
		3L-Encoder + Attention then Layer Dropout & 0.9 & 0.9 & \textbf{71.64} & \textbf{79.51} & 99.50 & 99.40 \\
		3L-Encoder + Layer then Attention Dropout & 0.8 & 0.6 & 71.22 & 78.66 & 99.45 & 99.44 \\
		3L-Encoder + Layer then Attention Dropout & 0.9 & 0.9 & 70.44 & 78.54 & 99.24 & 99.22 \\
		\hline\hline
	\end{tabular}
	}
	\label{tab1}
\end{table*}

\begin{table*}[ht]
	\centering
	\caption{Compared with Other SSL Methods, Phoneme and Speaker Classification Results on Librispeech, Accuracy (\%)}
	\scalebox{0.85} {
	\begin{tabular}{p{8.0cm}|p{2.3cm}|p{2.3cm}|p{2.3cm}|p{2.3cm}}
		\hline\hline
		\textbf{Pretraining Method} & \textbf{PhonemeLinear} & \textbf{Phoneme1Hidden} & \textbf{SpeakerFrame} & \textbf{SpeakerUtterance} \\
		\hline
		CPC~\cite{Oord2018Representation} & 64.6 & 72.5 & 97.4 & -- \\
		Modified CPC~\cite{Riviere2020Unsupervised} & 68.9 & -- & -- & -- \\
		AALBERT~\cite{Chi2020Audio} & -- & -- & 98.79 & 99.12 \\
		Mockingjay~\cite{Liu2020Mocking} & 64.3 & 76.8 & 68.4 & 96.1 \\
		TERA~\cite{liu2020tera} & 70.65~(65.1) & 78.51~(77.3) & \textbf{99.52}~(98.9) & \textbf{99.47}~(99.2) \\
		\hline
		\textbf{3L-Encoder + Attention then Layer Dropout (ours)} & \textbf{71.64} & \textbf{79.51} & 99.50 & 99.40 \\
		\hline\hline
	\end{tabular}
	}
	\label{tab2}
\end{table*}

\subsection{Layer Dropout}
For each transformer encoder layer $l$, the layer dropout method is applied on the output $\mathbf{X_l}$ with probability $p_{layer}$. Similar to attention dropout calculation, we firstly get the maximum absolute value $x_l^{max}$ of feature map $\mathbf{X_l}$ by spatial max-pooling:
\begin{equation}
x_l^{max} = \mathrm{MaxPool}(\vert\mathbf{X_l}\vert)
\end{equation}
Then, we design a binary masked map $\mathbf{M_l}$ to indicate whether each location $x_l^{ij}$ is dropped or not. Each element $m_l^{ij}$ of $\mathbf{M_l}$ is calculated as:
\begin{equation}
m_l^{ij} = \theta_{layer}(x_l^{ij}) = \begin{cases} 0, & \mbox{if}~\vert x_l^{ij}\vert>\lambda_{layer}x_l^{max} \\
1, & \mbox{otherwise}\end{cases}
\end{equation}
In which, $\lambda_{layer}\in[0,1]$ is the threshold ratio. $\vert\cdot\vert$ is the absolute value function, meaning that both positive and negative large value will be discarded. Finally, the binary masked map $\mathbf{M_l}$ is multiplied to original map $\mathbf{X_l}$, to get the final feature map:
\begin{equation}
\mathbf{X_l} = \mathbf{M_l}\odot\mathbf{X_l}
\end{equation}
where $\odot$ is denoted as element-wise matrix multiplication.



\section{Experimental Setup}
In this work, we focus on the representation extraction approach for downstream speech tasks. Following previous works, the experiments are in two stages: (1) pretrain the transformer encoder network by SSL, reconstructing the altered acoustic features, (2) extract the representations from the last layer of the model, and compare the performance on downstream tasks. In this section, we explored the experimental results of different dropout configurations on threshold ratio, and also visualized the changes of attention weight matrix and layer feature map by dropout regularization.

\subsection{Dataset}
For most experiments, we used publicly available LibriSpeech corpus~\cite{Panayotov2015Librispeech}. The \textit{train-clean-100} subset ($100$ hours) of LibriSpeech was used for pretraining. Like previous works of SSL, we used four downstream tasks for evaluation:
\begin{itemize}
\setlength{\itemsep}{0pt}
\setlength{\parsep}{0pt}
\setlength{\parskip}{0pt}
\item \textbf{PhonemeLinear}: phoneme classification with linear network
\item \textbf{Phoneme1Hidden}: phoneme classification with one hidden layer and linear network
\item \textbf{SpeakerFrame}: frame-wise speaker recognition
\item \textbf{SpeakerUtterance}: utterance-wise speaker recognition
\end{itemize}
For phoneme classification task, we used aligned phoneme labels and train/test split provided in the CPC~\cite{Oord2018Representation} and Modified CPC~\cite{Riviere2020Unsupervised}. Linear classifier and classifier with a single hidden layer are used to measure the linear separability of phonemes. For speaker recognition task, we also used the same train/test split as provided in the CPC. Two types of task, predicting speaker for each input frame and predicting speaker identity conditioning on averaged vector of each utterance, are provided.

\begin{figure*}[ht]
	\centering
	\subfigure[Original Weight Matrix] { \label{fig2a}
		\includegraphics[width=0.5\columnwidth]{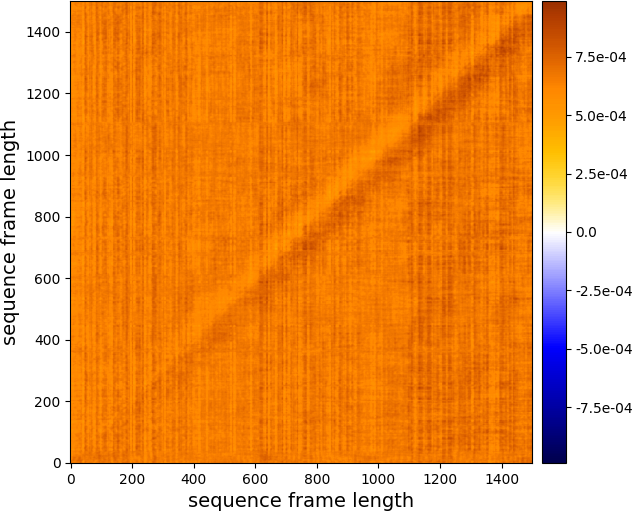}
	}
	\subfigure[After Attention Dropout] { \label{fig2b}
		\includegraphics[width=0.5\columnwidth]{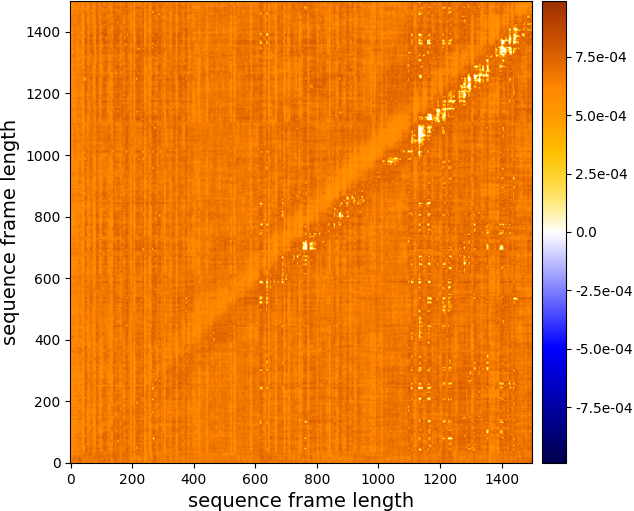}
	}
	\subfigure[Difference between (a) and (b)] { \label{fig2c}
		\includegraphics[width=0.5\columnwidth]{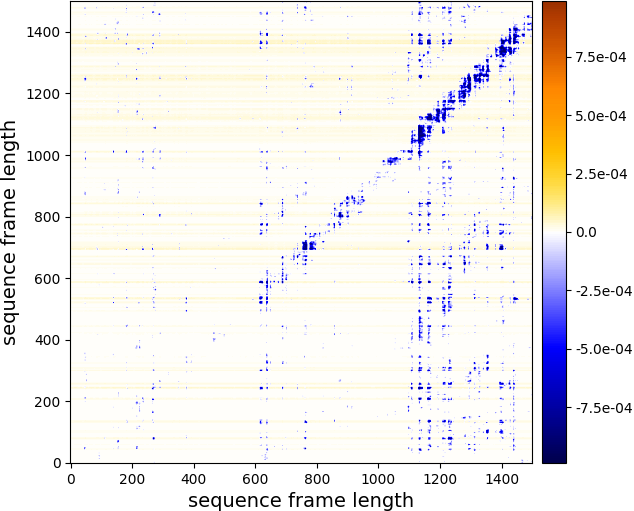}
	}
	\caption{Visualization of Attention Weight Matrix from Attention Dropout}
	\label{fig2}
\end{figure*}

\begin{figure*}[ht]
	\centering
	\subfigure[Original Feature Map] { \label{fig3a}
		\includegraphics[width=0.5\columnwidth]{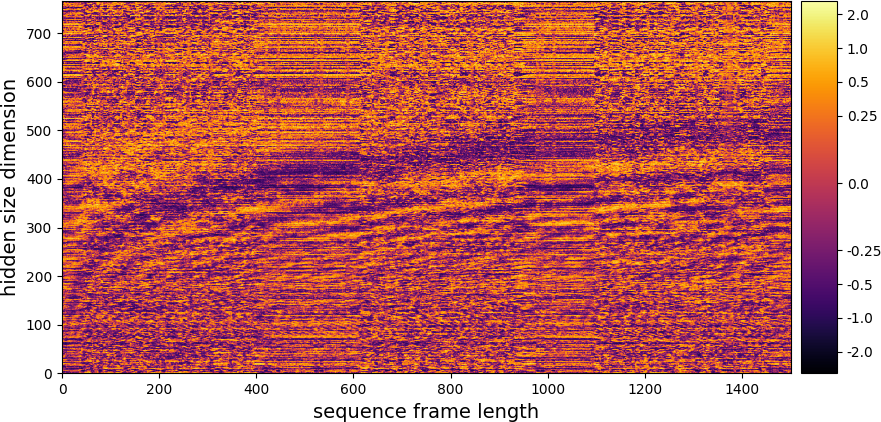}
	}
	\subfigure[After Layer Dropout] { \label{fig3b}
		\includegraphics[width=0.5\columnwidth]{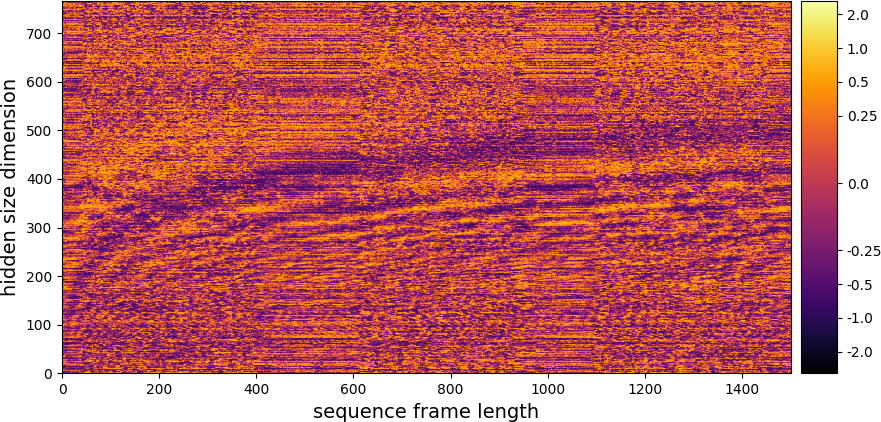}
	}
	\subfigure[Difference between (a) and (b)] { \label{fig3c}
		\includegraphics[width=0.5\columnwidth]{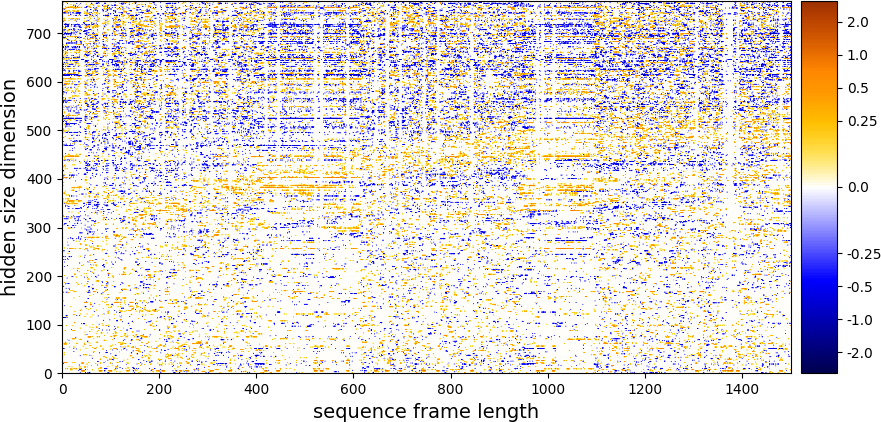}
	}
	\caption{Visualization of Layer Feature Map from Layer Dropout}
	\label{fig3}
\end{figure*}

\subsection{Configuration}
\label{Configuration}
We conducted all the experiments using the s3prl toolkit~\cite{S3PRL} on Pytorch framework. The parameters of self-supervised pretraining and downstream tasks are listed in Table~\ref{tab3}.

\begin{table}[ht]
	\centering
	\caption{Parameters of Pretraining and Downstream Tasks}
	\label{tab3}
	\scalebox{0.85} {
	\begin{tabular}{p{6.5cm}|p{2cm}}
		\hline\hline
		\multicolumn{2}{c}{\textbf{Self-Supervised Pretraining}} \\
		\hline
		input mel-scale features $D_{mel}$ & $80$ \\
		transformer encoder layers $L$ & $3$ \\
		attention hidden size $D_{attn}$ & $768$ \\
		multi-heads $H$ & $12$ \\
		feed-forward dimension & $3072$ \\
		attention dropout probability $p_{attn}$ & $0.1$ \\
		layer dropout probability $p_{layer}$ & $0.1$ \\
		batch size & $32$ \\
		training steps & $200k$ \\
		\hline\hline
		\multicolumn{2}{c}{\textbf{Phoneme Classification Task}} \\
		\hline
		phoneme classes & $41$ \\
		one hidden layer dimension & $768$ \\
		batch size & $32$ \\
		training steps & $20k$ \\
		\hline\hline
		\multicolumn{2}{c}{\textbf{Speaker Recognition Task}} \\
		\hline
		speaker classes & $251$ \\
		batch size & $32$ \\
		training steps & $20k$ \\
		\hline\hline
	\end{tabular}
    }
\end{table}

The overall architecture is three-layers transformer encoder network. The input audio is encoded with $80$ mel-scale features. Each transformer encoder layer contains two parts: (1) self-attention layer ($768$ dimension and $12$ multi-heads) with attention dropout ($10\%$ probability), (2) feed-forward layer ($3072$ dimension) with layer dropout ($10\%$ probability). The models were pretrained by total $200k$ steps with batch size $32$.

For phoneme classification task, we adopt the common setup using $41$ possible phoneme classes, and $768$ dimension for one hidden layer. For speaker recognition task, the dataset consists of $251$ speakers. Besides, we trained all the downstream tasks by $20k$ steps. The parameters of the pretrained models are frozen, when the downstream tasks are trained.

\subsection{Results}
We conducted the experiments on different configurations of attention threshold ratio $\lambda_{attn}$ and layer threshold ratio $\lambda_{layer}$. As shown in Table~\ref{tab1}, we found that the three-layers transformer encoder model achieves best performance with $\lambda_{attn}=0.8$ for attention dropout and $\lambda_{layer}=0.6$ for layer dropout. The threshold cannot be set too small, otherwise too much high activation regions will be discarded and the performance will degrade. In addition, the closer threshold is to $1.0$, the closer results are to 3L-TERA-base~\cite{liu2020tera}. For fair comparison, all of the experimental results in Table~\ref{tab1} were performed on the same configurations in Table~\ref{tab3}, and we referenced the numbers of TERA in the $(\cdot)$.

We also investigated three fusion strategies of two dropout regularization, (1) Attention \& Layer Dropout, conducting two dropout together with half dropout probability $0.05$, (2) Attention then Layer Dropout, pretraining $100k$ steps with attention dropout, then another $100k$ steps with layer dropout, (3) Layer then Attention Dropout. In our experiments, we found Attention then Layer Dropout with threshold ratio $0.9$ works better than two other fusion strategies, and outperforms the method of attention or layer dropout alone as presented in Table~\ref{tab1}.

As depicted in Table~\ref{tab2}, we compared our approach with other SSL methods. We choosed the published results using the same training set, \textit{train-clean-100} of LibriSpeech. Our best model (Attention then Layer Dropout) achieves $1.40\%$ relative improvement on the accuracy of PhonemeLinear task, and $1.27\%$ of Phoneme1Hidden task, over the original TERA-base model. Despite the results of speaker recognition tasks are very close with each other, our approach outperforms most of the listed methods on the downstream tasks.

\subsection{Visualization}
In Figure~\ref{fig2} and Figure~\ref{fig3}, we visualized the attention weight matrix from attention dropout and layer feature map from layer dropout. After the attention dropout, the most nearby attention weights of each location in Figure~\ref{fig2a} are discarded (see Figure~\ref{fig2b}). The rest attention weights are distributed to far distant locations (see yellow lines in Figure~\ref{fig2c}). By contrast, the layer dropout prefers to function as regularization. The layer dropout will suppress the most negative activations (see yellow regions in Figure~\ref{fig3c}) and discard largest positive values (see blue regions in Figure~\ref{fig3c}). As a result, the feature map (Figure~\ref{fig3b}) becomes smoother than the original one (Figure~\ref{fig3a}). Overall, the visualization demonstrates that with dropout regularization, the model suppresses the overemphasized local features and captures more global information.

\section{Conclusions}
In this paper, we proposed to use attention dropout and layer dropout in the SSL of speech representation. Attention dropout reweights the multi-head attention matrix of each transformer encoder layer. Layer dropout discards the most discriminative activation regions by spatial max-pooling. The experiments show that downstream phoneme classification and speaker recognition tasks can obtain substantial performance improvements with attention and layer dropout. In future works, we will explore the effect of dropout on other downstream tasks like speech recognition. We are also interested to investigate the performance of dropout regularization on various SSL models besides the transformer encoder architecture.

\section{Acknowledgement}
\label{sec:ack}
This paper is supported by National Key Research and Development Program of China under grant No. 2018YFB0204403 , No. 2017YFB1401202 and No. 2018YFB1003500. Corresponding author is Jianzong Wang from Ping An Technology (Shenzhen) Co., Ltd.

\clearpage
\bibliographystyle{IEEEtran}

\bibliography{IS2021_Unsupervised_Dropout}

\end{document}